\documentstyle{article}

\include{psfig}
\newcommand{\newc}{\newcommand}

\newc{\beq}{\begin{equation}}
\newc{\eeq}{\end{equation}}
\newc{\beqa}{\begin{eqnarray}}
\newc{\eeqar}{\end{eqnarray}}
\newc{\beqar}{\begin{eqnarray}}
\newc{\eeqa}{\end{eqnarray}}
\newcommand{\bd}{\begin{displaymath}}
\newcommand{\ed}{\end{displaymath}}
\newc{\pa}{\partial}
\newc{\bom}{\boldmath}
\newc{\btd}{\bigtriangledown}
\newc{\rarrow}{\rightarrow}
\newc{\ep}{\epsilon}
\newc{\beqas}{\begin{eqnarray*}}
\newc{\eeqas}{\end{eqnarray*}}
\newc{\mb}{\mbox}
\newc{\tm}{\times}
\newc{\hu}{\hat{u}}
\newc{\hv}{\hat{v}}
\newc{\hk}{\hat{K}}
\newc{\ld}{\lambda}
\newc{\sg}{\sigma}
\newc{\p}{\psi}
\newc{\kt}{\rangle}
\newc{\br}{\langle}
\newc{\ra}{\Rightarrow}
\newc{\dg}{\dagger}
\newc{\non}{\nonumber}
\newc{\ul}{\underline}
\newc{\longra}{\longrightarrow}
\newc{\hs}{\hspace}
\newc{\eps}{\epsilon}
\newc{\longla}{\longleftarrow}
\newc{\bmyquote}{\begin{quote} \begin{footnotesize} \begin{sf}}
\newc{\emyquote}{\end{sf} \end{footnotesize} \end{quote} }
\begin{document}
\def\thesection{\Roman{section}.}
\def\thesubsection{\Alph{subsection}.}

\begin{center}

\bigskip

{\large \bf A Bethe Ansatz Study of Free Energy and Excitation Spectrum for even spin Fateev Zamolodchikov Model} 
\bigskip

{\large Subhankar Ray} {\scriptsize $^{^{\dag }}$} {\scriptsize $^{^{*}}$}
\footnote {email: subho@juphys.ernet.in},
{\large J. Shamanna} {\scriptsize $^{^{\ddag}}$}
\bigskip

{\scriptsize $^{^{\dag }}$} 
Department of Physics, Jadavpur University, Calcutta 700 032, India. \\
{\scriptsize $^{^{*}}$}
C. N. Yang Institute for Theoretical Physics, Stony Brook, NY 11794, USA. \\
{\scriptsize $^{^{\ddag }}$}
Department of Physics, Visva Bharati University, Santiniketan 731235, India.
\bigskip

\end{center}

\begin{abstract}
\noindent
A Bethe Ansatz study of a self dual $Z_N$ spin model is
undertaken for even spin system.
One has to solve a coupled system of Bethe Ansatz Equations (BAE) 
involving zeroes of
two families of transfer matrices.
A numerical study on finite size lattices is done for identification 
of elementary excitations over the Ferromagnetic and Antiferromagnetic
ground states.
The free energies for both Ferromagnetic and Antiferromagnetic ground states 
and dispersion relation for elementary excitations are found.
\end{abstract}

%PACS: 05.50.+q; 75.10.Jm; 02.90.+p; 11.25.Hf

\setcounter{equation}{0}

\section{Introduction}
The present model was first proposed 
in 1982 by V.A. Fateev and 
A.B. Zamolodchikov \cite{pFZ1} as 
a two dimensional self dual $Z_N$ lattice model with nearest neighbor
spin-spin interaction.  
Baxter, Bazhanov and Perk \cite{pBBP} discovered a set of 
functional equations involving families of Chiral Potts (CP)
transfer matrices. Fateev Zamolodchikov model (FZM) was shown to be
a non-chiral self dual limit of Chiral Potts \cite{Sray01,srjmp}.
The Chiral Potts transfer matrix functional equations were used
to obtain transcendental equations (Bethe Ansatz Equations) 
\cite{pbethe} for the zeroes of the transfer matrices
for the present problem \cite{Sray01,srjmp}.
In the general FZM model the Bethe Ansatz Equations (BAE) are coupled
involving two automorphically connected families of transfer matrices as in the CP case \cite{pBBP}.
In the odd spin case these families are connected by simple transformations
and the BAEs greatly simplify and decouple requiring us to solve only 
one set of equations \cite{srjmp}.
For this odd spin case alone, Albertini obtained the 
ferromagnetic ground state \cite{albert1}. 
A unified treatment for the ground state 
of odd and even spin FZM can be found in a previous work \cite{srjmp}.
For $N=4$ FZM a comprehensive study has been done. This includes
determination of exact energy values 
and central charge \cite{srpl1}, and
completeness and classification of Bethe states 
\cite{srpl2}.
The present work demonstrates the
study of these Bethe Equations in both ferromagnetic and 
antiferromagnetic cases for finding the free energy and 
elementary excitations. \\

In a generic situation for Fateev Zamolodchikov model (to be specific 
we shall use the even spin case) we obtained coupled Bethe Ansatz equations,
\beqar
\prod_{k=1}^{L_{Uq}} \frac{\sinh (\lambda_j -\bar{\lambda}_k - i \gamma)}{\sinh (\lambda_j -\bar{\lambda}_k + i \gamma)} &=& (-1)^{M+1} \left[ \frac{\sinh 2(\lambda_j + i s \gamma)}{\sinh 2(\lambda_j - i s \gamma)} \right]^{2M} \\
\prod_{k=1}^{L_{q}} \frac{\sinh (\bar{\lambda}_j -\lambda_k - i \gamma)}{\sinh (\bar{\lambda}_j -\lambda_k + i \gamma)} &=& (-1)^{M+1} 
\eeqar
where 
$\{\lambda_j\}$ and $\{\bar{\lambda}_j\}$ are the spectral variables
for transfer matrices $T_q$ and $T_{U_q}$ respectively 
and $\gamma=\frac{\pi}{2N}$  and  $s=\frac{1}{2}$.

The first equation is quite similar to the generic case of
Bethe Equation. However it involves zeroes of two different
transfer matrices $( \lambda,\; \bar{\lambda} )$ that are coupled further 
by a second equation whose form is quite unique.

The generic case of most commonly encountered Bethe Ansatz Equation looks like

\bd
\prod_{k=1}^{L}\frac{\sinh(\lambda_{j}-\lambda_{k}-i\gamma)}{\sinh(\lambda_{j}-\lambda_{k}+i\gamma)}=(-)^{M+1}\left[ \frac{\sinh(\lambda_{j}-i S \gamma)}{\sinh(\lambda_{j}+i S \gamma)} \right]^{2M}
\ed
where $S$ is the spin and $\gamma$ is the anisotropy parameter
of the model. In Bethe's original paper \cite{pbethe}, he studied the
case $\gamma \rightarrow \infty$, where the hyperbolic functions
reduce to rational ones. Note that in the generic case we have 
only one type of $\lambda_{j}$ as opposed to two species of
$\lambda_j$ and $\bar{\lambda}_j$ as they appear in the present problem.  \\

The standard procedure of calculating the physical 
quantities, e.g., the energy spectrum, dispersion curves,
free energy, is to assume that the solutions for the Bethe 
Equation are given by the string hypothesis.
Starting with the work of Bethe there has been a great deal of study in 
these complex solutions of BAE.
They appear in the form,

\bd
\lambda_{\alpha ,k}^{(n,\nu)}=\lambda_{\alpha}^{(n,\nu)}+\frac{\gamma}{2}(n+1-2k)i+\frac{(1-\nu)\pi}{4}i+\delta_{\alpha ,k}^{(n,\nu)}\ \ \ k=1,2 \ldots n
\ed
where $\lambda_{\alpha}^{(n,\nu)}$ is the real part, $n$ is its length, 
$k$ runs from $0$ through $n$ labelling the root. 
The coefficient $\nu$ takes on the value $(+1)$ (positive parity) or $(-1)$
(negative parity); $\delta_{\alpha ,k}^{(n,\nu)}$ vanishes regularly as $M \rightarrow \infty$. \\

However Bethe himself realized that for large $M$ not all solutions of BAE
are of the form (above) with $lim_{M\rightarrow \infty} Im (\delta )=0$.
Modern work for the case of $lim_{M\rightarrow \infty} Im (\delta )\neq 0$
was initiated by Destri and Lowenstein  and Woynarovich who
introduced the definition of narrow pairs for $lim_{M\rightarrow \infty} Im (\delta ) < 0$
and wide pairs for $lim_{M\rightarrow \infty} Im (\delta ) > 0$
and was furthered by Avdeev and D$\ddot{o}$rfel \cite{AD,BdVV} who introduced 3 classes
for $lim_{M\rightarrow \infty} Im (\delta )$.
It is clear that none of the existing approaches to BAE for $S$ integer or
half integer is sufficiently refined to answer the reality or completeness
question for the Hamiltonian.  \\

In the general spin hypothesis framework one obtains equations for the centers of the
strings by multiplying the Bethe Ansatz equations over different
members of the same string and then taking the logarithm of the
resulting equation. This yields

\bd
\frac{1}{2\pi}\Theta_{j}^{(1)}(\lambda_{\alpha}^{j})-\frac{1}{2\pi M}\sum_{k}\sum_{\beta =
1}^{M^{(k)}}\Theta_{jk}^{(2)}(\lambda_{\alpha}^{(j)}-\lambda_{\beta}^{(k)})=\frac{I_{\alpha}^{(j)}}{M}
\ed
where $M^{(k)}$ is the number of $k$-strings and 
\bd
\Theta_{j}^{(1)}(\lambda)=2\sum_{\ell =1}^{n_{j}}\phi(\lambda,n_{j}+2s-2\ell +1,\nu_{j}) 
\ed
\bd
\Theta_{jk}(\lambda)=\phi(\lambda,n_{j}+n_{k},\nu_{j}\nu_{k})+\phi(\lambda,|n_{j}-n_{k}|,\nu_{j}\nu_{k})+\!\!\!\!\!\!\! \sum_{\ell =1}^{{\rm min}(n_{j},n_{k})-1}\!\!\!\!\!\!\! 2\phi(\lambda,|n_{j}-n_{k}|+2\ell,\nu_{j} \nu_{k})
\ed
and 
\bd
\phi(\lambda,n,\nu)=\left\{
\begin{array}{l}
2 \nu \arctan(\cot(\frac{n\gamma}{2})^{\nu}\tanh(\lambda))\\
\mbox{$0$ if $ n\gamma=q\pi \ \ q\in Z$}
\end{array}
\right.
\ed
In certain cases, e.g., $\delta$-function Bose gas, it can be proved that the 
solutions are real, and no such multiplying of string components is necessary
\cite{pkorebook}. In such cases, the integers designating the branches of 
logarithms 
uniquely characterize the states and may be viewed as quantum numbers for the
states. A monotonic relation is shown to exist between the integers and the
values of the spectral variable $\lambda_{\alpha}^{(n,v)}$.
In almost all subsequent work in the field this unique characterization of
states by integers and their monotonic relation to solutions have been 
assumed. In few cases some counting argument is attempted 
\cite{pbethe,ptak71,pkiri85,pkiri87,pleon82} to justify this assumption. 
In the $N \rightarrow \infty$ limit, after introducing the concept 
of density of string centers, one obtains a coupled set of 
integral equations.
These equations are manipulated to calculate the energies of
the ground state and low lying excited states. 
For the spin 4 system, complete classification of Bethe states,
and exact calculation for finite and infinite systems is already known
\cite{srpl1, srpl2}.
 
In the present problem, we had to deal with a doubly coupled set of
integral equations; two sets of coupled equations both involving zeroes
of two types of transfer matrices. However linearity of the equations made it
possible to solve for the ground state and elementary excitations 
by Fourier transform method. 
A study of the classification of roots is undertaken.
However one has to keep track of the added complexity of 
having to handle $T_q$ and $T_{U_q}$  simultaneously.
One can identify the ground state and elementary excitations on the basis 
of this numerical
study. The excitation spectrum and dispersion relations can hence be
calculated. \\

%%%%%%%%%%%%%%%%%%%%%%%%%%%%%%%%%%%%%%%%%%%
% 	Now begins description of the model 
%   Additional material added on referee advice (ref JMP)
%%%%%%%%%%%%%%%%%%%%%%%%%%%%%%%%%%%%%%%%%%%%	

\section{Fateev-Zamolodchikov model}
V.A. Fateev and A.B. Zamolodchikov proposed in 1982 
a two dimensional self dual $Z_N$ lattice spin model with nearest neighbor 
interaction. They obtained this model as the self dual \cite{pKC}
solution of the star-triangle relations \cite{Bbook}.

A general $Z_N$ model can be defined as follows. On a two dimensional
rectangular lattice the lattice sites are occupied by a spin variable
$z$ which takes its values in the group $Z_N$ $[z^N=1]$. If one designates
the sites on the lattice by a two dimensional integer-valued vector $\boldmath x$,
one can write down the partition function of the statistical $Z_N$ model with
nearest neighbor interaction as:

\beq
Z = \sum_{\{z\}} \prod_{\mbox{\boldmath $x$}} \prod_{\sigma = \pm} w^{(\sigma)}(z(\mbox{\boldmath $x$}), z(\mbox{\boldmath $x$}+\mbox{\boldmath $\epsilon$}_{\delta})).  \\
\eeq

where the sum runs over all values of the variable $z$ in every site of the lattice. 
The functions $w^{\sigma},\;\; (\sigma = \; \pm 1)$  are the weight functions 
corresponding to the interaction between spins on the neighboring sites of the lattice 
in horizontal $(\sigma = 1)$ and vertical $(\sigma = -1)$ directions respectively. 
The vectors $\boldmath \epsilon_1 = (1,0)$ and $\boldmath \epsilon_{-1} = (0,1)$ are 
the basis vectors of the lattice. \\

\begin{figure}
\centerline{\psfig{figure=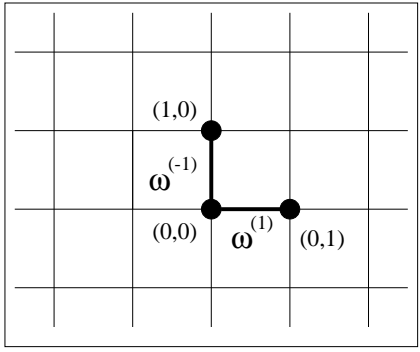,height=2in}}
\vspace{0.in}
\caption{Fateev Zamolodchikov Model on a square lattice}
\label{fzmfig.ps}
\end{figure}

In the absence of external fields, the most general interaction between two neighboring
spins after appropriate normalization is given by
\beq
w^{(\sigma)}(z_1,z_2) = 1 + \sum_{i=1}^{N-1} x_i^{(\sigma)} \cdot (z_1 z_2^{\star})^i
\eeq
where superscript $^*$ denotes complex conjugate.
Reality of $w^{(\sigma)}(z_1,z_2)$ imposes on the parameters the following restriction
\beq\label{eq5}
x_i^{(\sigma)} = x_{N-i}^{(\sigma)}
\eeq
The dual transformation of the statistical weights are given by
\beq
\tilde{x}_i^{(\sigma)} = \left( 1 + \sum_{k=1}^{N-1} x_k^{(-\sigma)} 
\omega^{k i} \right) \cdot \left(1 + \sum_{k=1}^{N-1} x_k^{(-\sigma)}
\right)^{-1}
\eeq
where $\omega = exp(2 \pi i/N)$. The region of self duality is then given by
\beq\label{eq7}
\tilde{x}_i^{(\sigma)} = x_i^{(\sigma)}  \\
\eeq
Let the parameters $x_i^{(\sigma)}$ be represented by a family of functions
$W_i(\alpha)$ of auxiliary parameter $\alpha \; \in {\cal C}$
\beq
x_i^{(1)} = W_i(\alpha) , \;\;\;\;  x_i^{(-1)} = W_i(\pi - \alpha).
\eeq
 The star-triangle relation \cite{CoBook,Za2} on $x_i(\alpha)$ :
\beqar\label{eq9}
\sum_{k=0}^{N-1} W_{n_1-k} (\alpha) W_{n_2-k} (\pi-\alpha -\alpha') W_{n_3-k} (\alpha') = \nonumber \\
c(\alpha,\alpha') W_{n_2-n_3} (\pi-\alpha) W_{n_1-n_3} (\alpha + \alpha') W_{n_1-n_2} (\pi-\alpha')
\eeqar
\noindent
The particular solution of Eq. (\ref{eq9}) that possesses the self duality 
property, e.g. Eq. (\ref{eq7}), is given by:
\beq\label{eq10}
W_0 = 1, \;\;\; W_n(\alpha) = \prod_{k=0}^{n-1} \frac{\sin[\pi k/N + \alpha / 2N]}{\sin[\pi (k+1)/N - \alpha /2N]}.
\eeq
Denoting $x_n^{(1)} = W (n\mid u)$ and $x_n^{(-1)} = \overline{W} (n\mid u)$ we get
\beq\label{eq11}
\frac{W (n|u)}{W (0|u)}=\prod_{j=1}^{n}\frac{\sin(\pi j/N-\pi/2N-u)}{\sin(\pi j/N-\pi/2N+u)}
\eeq
\beq\label{eq12}
\frac{\overline{W} (n|u)}{\overline{W} (0|u)}=\prod_{j=1}^{n}\frac{\sin(\pi j/N-\pi/N+u)}{\sin(\pi j/N-u)} \\
\eeq

We adopt the normalization $W (0|u)=\overline{W} (0|u)=1$. The ``physical region'' defined by
non-negative real Boltzmann Weights (BW) , is given by $u \in [0,\pi/2N[$. 
For $N=2,3$ Eq. (\ref{eq11}) and 
Eq. (\ref{eq12}) simply
reduce to the self-dual critical Potts model. For $N=4$ it gives a particular case of
critical Ashkin-Teller model. Fateev and Zamolodchikov propose that for $N=5,7$ the solution
describes the critical bifurcation points in the phase diagram of Alcaraz and Koberle \cite{AK1}. \\

\section{Chiral Potts model and connection to FZM}
On the sites of a two dimensional lattice of size $\cal{M} \times \cal{N}$
denoted by two dimensional vector
$(j,k)$ with integer entries, we place $Z_N$ spins $\sigma_{j,k}$.
The spins $\sigma_{j,k}$ are classical variables satisfying:
\bd
\sigma_{j,k}^N = 1
\ed
i.e., $\sigma_{j,k} = \omega^\nu, \nu \in \{0,1,2,\dots N-1\}$ where $\omega$ is
the complex $N^{th}$ root of unity with the minimum argument.
\bd
\omega = e^{2 \pi i/N}
\ed

The energy corresponding to a given configuration of spins $\{ \sigma_{j,k} \}$ is
\bd
{\cal E} = - \sum_{\{j,k\}} \sum_{n=1}^{N-1}\{ E_n^h \cdot (\sigma_{j,k} \sigma_{j,k+1}^{\star})^n + E_n^v \cdot (\sigma_{j,k} \sigma_{j+1,k}^{\star})^n\}.
\ed
Row index $j$ runs over $1$ to $\cal M$ and column index $k$ runs over 
$1$ to $\cal N$ with periodic boundary condition in both directions implied. 

\begin{figure}
\centerline{\psfig{figure=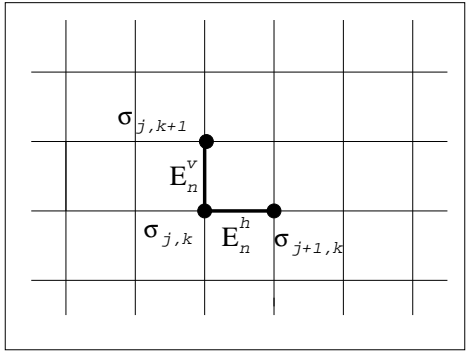,height=2in}}
\vspace{0.in}
\caption{Square lattice Chiral Potts Model}
\label{sqrl.ps} 
\end{figure}

In Chiral Potts a subspace of the 
coupling parameters $(E_n^h,E_n^v)$ is chosen which has a built-in handedness, 
or phase. The energy of a nearest neighbor pair is chosen as
\bd
{\cal E}_{pair}^{h,v}(\sigma_1,\sigma_2) = - \sum_{n=1}^{N-1} E_n^{h,v} \cdot (\sigma_1 \sigma_2^{\star})^n
\ed
where $E_n^{h,v} = |E_n^{h,v}| \cdot e^{i \delta_n}$. The local Boltzmann 
weights (BW) can now be easily defined as:
\bd
W^{h,v}(n) = e^{\frac{1}{k_BT} \sum_{j=1}^{N-1} E_j^{h,v} \omega^{jn}}
\ed
Let us denote two adjacent row configurations by $\{l\}$ and $\{l'\}$ ( $l$ corresponds
to the lower row), where
\bd
\{l\} = \{ \omega^{l_j} \mid j=1(1){\cal N} \;\; and \;\; l_j \in {0,1,\dots,N-1} \}.
\ed

\noindent
The row to row transfer matrix is given by:
\bd
T_{\{l\},\{l'\}} = \prod_{j=1}^{\cal N} W^h(l_j - l'_{j+1}) \cdot W^v(l_j - l'_j).
\ed

\begin{figure}
\centerline{\psfig{figure=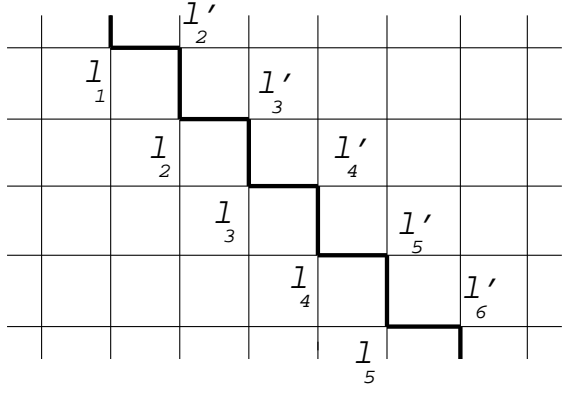,height=2in}}
\vspace{0.in}
\caption{Transfer Matrix $T_{\{l\},\{l'\}}$}
\label{matrix.ps}
\end{figure}
It has been shown by several authors \cite{pFZ1,YMPTY,MPTS,BPY} that the transfer 
matrices corresponding to the interaction parameters belonging to the Chiral Potts 
submanifold commute.
\bd
[ T, T' ] = 0
\ed

The self dual Chiral Potts model is given by BWs
\beq\label{eq13}
\frac{W_{pq}(n)}{W_{pq}(0)}=\prod_{j=1}^{n}\frac{b_{q}-\omega^{j}a_{p}}{b_{p}-\omega^{j}a_{q}}.
\eeq
\beq\label{eq14}
\frac{\overline{W}_{pq}(n)}{\overline{W}_{pq}(0)}=\prod_{j=1}^{n}\frac{\omega a_{p}-\omega^{j}a_{q}}{b_{q}-\omega^{j}b_{p}}.
\eeq
where $\omega=\exp(2\pi i/N)$ and the paired complex variables $(a,b) \; \in \; {\cal C}^2$ 
satisfy the constraint
\beq\label{eq15}
a_{x}^{N}+b_{x}^{N}=\kappa
\eeq
$\kappa \in [0,1]$, and $x = p$ or $q$. In the non chiral limit, when $\kappa=0$, we can 
parametrize $(a_x,b_x)$ in Eq. (\ref{eq15}) as:
\beq
a_{x}=e^{2ix}\ \ \  b_{x}=\omega^{1/2}e^{2ix}.
\eeq
Defining $u=q-p$ Eq. (\ref{eq13}) and Eq. (\ref{eq14}) reduce to 
Eq. (\ref{eq11}) and Eq. (\ref{eq12}).
However we retain suffixes $(p,q)$ in the Boltzmann weights $W_{pq} (n |u)$
and $\overline{W}_{pq} (n |u)$ to signify that these BWs are obtained from
the Chiral Potts BWs defined in terms of $p$ and $q$ variables.

The transfer matrix for the FZM can be constructed from the BWs as:
\beq
T_{p,q}^{\mbox{\boldmath $n,n'$}}(u) = \prod_{k=1}^{M}\overline{W}_{pq} (n_{k}-n_{k}'|u) W_{pq} (n_{k}-n_{k+1}'|u).
\eeq
where $M$ is the number of sites in each row and periodic boundary condition
is implied. These transfer matrices for different spectral variable $u$ form a 
commuting set. This can be argued from 
the fact that these transfer matrices come as a limit of CP transfer matrix, which are
known to be commuting. A more direct argument would be that Fateev and
Zamolodchikov obtained FZM BWs as solutions of star-triangle relation, and hence
the transfer matrix constructed out of them ought to commute.

\beq
[T(u),T(u')]=0 \ \ \   \forall \;\;u,u' \in C
\eeq

Transfer matrix $T(u)$ reduces to identity operator for $u \rightarrow 0$. An 
expansion of $T(u)$ gives us the associated spin chain Hamiltonian $H$.
\beq\label{eq19}
T(u)=1-Mu\sum_{n=1}^{N-1}\frac{1}{\sin(n\pi/N)}-uH+O(u^{2})
\eeq
\beq\label{eq20}
H=-\sum_{k=1}^{M}\sum_{n=1}^{N-1}\frac{1}{\sin(n\pi/N)}(X_{k}^{n}+Z_{k}^{n}Z_{k+1}^{-n}).
\eeq
where $X$ and $Z$ are defined as
\begin{eqnarray*}
X_{k}|n_{1} \dots n_{k} \dots n_{M}> & = &  |n_{1} \dots n_{k}+1 \dots n_{M}> {\rm mod} N\\
Z_{k}|n_{1} \dots n_{k} \dots n_{M}> & = & \omega^{n_{k}}|n_{1}\dots n_{k}\dots n_{M}>  \;\; .  \\
\end{eqnarray*}

Eq. (\ref{eq19}) and Eq. (\ref{eq20}) imply that each Hamiltonian 
commutes with all the
transfer matrices and their associated Hamiltonians. Thus it has an infinite 
set of conserved charges in involution. However only a subset of them, whose 
number is equal to the degrees of freedom of the system, are independent.

In order to obtain the zeros of the eigenvalues of the transfer matrix $T_q$, 
we will use functional equations connecting $T_q$ with its automorphically 
conjugate partners. Thus it is important to understand the relevant 
automorphisms of the constraint Eq. (\ref{eq15}).
It has been claimed in the previous section that
the transfer matrices constructed out of CP BWs, Eq. (\ref{eq13}) and 
Eq. (\ref{eq14}).
commute as long as they satisfy Eq. (\ref{eq15}).
For any $(a,b) \in {\cal C}^2$ satisfying the above relations there exist other 
complex pairs connected to them which satisfy the same relation. Two such automorphic 
relations of importance are,
\beq
R(a, b) = (b, \omega a) , \hskip 1.5cm
U(a, b) = (\omega a, b)
\eeq
It is rather straightforward to check;
\bd
a_{Rx}^N + b_{Rx}^N = \kappa, \hskip 1cm a_{Ux}^N + b_{Ux}^N = \kappa
\ed
from the relation 
\bd
a_x^N + b_x^N = \kappa
\ed

If one makes an attempt to go over from CP BWs to FZM BWs through a
limiting process, one gets the following relations for $W_{pRq}$,  
$\overline{W}_{pRq}$, $W_{pUq}$ and $\overline{W}_{pUq}$,
\beqa
\frac{W_{pRq}(n|u)}{W_{pRq}(0|u)}&=&\prod_{k=1}^{n}\frac{\sin(\pi k/N -\pi /N -u)}{\sin(\pi k/N +u)} \\
\frac{\overline{W}_{pRq}(n|u)}{\overline{W}_{pRq}(0|u)}&=& \prod_{k=1}^{n}
\frac{\sin(\pi k/N -\pi /2N +u)}{\sin(\pi k/N -\pi /2N -u)} \\
\frac{W_{pUq}(n|u)}{W_{pUq}(0|u)}&=& e^{-\frac{\imath \pi n}{N}}\prod_{k=1}^{n}\frac{\sin(\pi k/N -\pi /2N -u)}{\sin(\pi k/N +\pi /2N +u)} \\
\frac{\overline{W}_{pUq}(n|u)}{\overline{W}_{pUq}(0|u)}&=& e^{\frac{\imath \pi n}{N}}\prod_{k=1}^{n}\frac{\sin(\pi k/N +u)}{\sin(\pi k/N -u)}
\eeqa
Thus in the non chiral limit, $T_q \longrightarrow T_q (u)$ and 
$T_{Rq}\longrightarrow T_q (u+\pi/2N)$. There is no simple relation between $T_q$
and $T_{Uq}$ though. However, we do feel that there must exist
some nontrivial mapping between $T_q$ and $T_{Uq}$ whose understanding will
unravel the connection between the zeroes of $T_q$ and $T_{U_q}$ and will
give the satisfactory derivation of completeness of states.

\section{Bethe Ansatz type equations for the even $N$ FZM}

We define the normalized transfer matrices by removing their denominators,
\bd
T_q^{N}(u)=[g_q(u)\overline{g}_q(u)]^{M} T_q(u)
\ed
where
\bd
g_q (u)=\prod_{j=1}^{N/2}\sin(\frac{\pi j}{N}-\frac{\pi}{2N}+u) \ \ \ \
\overline{g}_q (u)=\prod_{j=1}^{N/2}\sin(\frac{\pi j}{N}-u)
\ed
One must note that the superscript in $T_q^N$ denotes ``normalize'' and is not
related to the spin quantum number $N$.
Each entry of $T_q^{N}(u)$ is a product of $NM$ sines and it has the general form
\bd
\prod_{k=1}^{NM}(c_{k}^{(1)}e^{iu}+c_{k}^{(2)}e^{-iu})
\ed

The calculation of this section goes in the same spirit as that of odd $N$ case.
Hence we only quote the results.
\begin{eqnarray}\label{e15}
\Lambda_{Q=0}(u)&=& \left[ \frac{g_q (0)\overline{g}_q (0)}{g_q (u)\overline{g}_q (u)} \right]^{M}
\prod_{k=1}^{L}\frac{\sin(u-v_{k})}{\sin v_{k}} \\ \nonumber \\
L&=&A+B=N M \nonumber
\end{eqnarray}
The normalization has been fixed by $T_q (0)=1_{id}$.

The momentum $(P)$ is given by,
\beq
e^{i P} = \Lambda_{Q}(u=\frac{\pi}{2N})= 
\left[ \frac{g_q (0)\overline{g}_q (0)}{g_q (\frac{\pi}{2N})
\overline{g}_q (\frac{\pi}{2N})} \right]^{M} 
\prod_{k=1}^{L}\frac{\sin(\frac{\pi}{2N}-v_{k})}{\sin v_{k}} 
\eeq

Now we turn to the sectors $Q \neq 0$, and the symmetry under charge conjugation allows us to consider the sectors $Q=1,2,\ldots,(N-1)/2$ only. While we have not been able to obtain a proof like the one given above, one can show that, in the sector $Q$
\begin{eqnarray*}
a)\ \ A,B &\leq& \frac{N M}{2} -Q, \hskip 1cm Q=1,2\dots \frac{N}{2} \\
b)\ \ A,B &\geq& \frac{N M}{2} - \frac{N}{2}
\end{eqnarray*}

Following similar argument as before, we arrive at
\bd
A=B= \frac{N M}{2} - Q
\ed
The reader must be warned that this conclusion lacks rigor just like in the case 
of $N$ odd. The factorization in terms of sines can be carried out 
without the appearance of a phase $(e^{2iu})^{\pm (B-A)}$. 
We assume this to be true also for the others $Q$ sectors, and arrive 
at the general form
\begin{eqnarray}\label{e16}
&& \Lambda_{Q}(u)= \left[ \frac{g_q (0) \overline{g}_q (0)}{g_q (u) \overline{g}_q (u)} \right]^{M}
\prod_{k=1}^{L}\frac{\sin(u-v_{k})}{\sin v_{k}}\\ \nonumber \\
&& L=N M-2Q, \hskip 1cm Q=0,1,\dots \frac{N}{2}, \hskip 1cm 
\Lambda_{N-Q}(u)=\Lambda_{Q}(u)\nonumber
\end{eqnarray}
From this, the eigenvalue of $H$ is easily found to be
\begin{equation}\label{e17}
E=\sum_{k=1}^{L}\cot v_{k}\ \ -2M \! \sum_{j=1}^{N/2} \! 
\cot (\pi j/N)
\end{equation}
The momentum $(P)$ is given by,
\beq
e^{i P} = \Lambda_{Q}(u=\frac{\pi}{2N}) =
\left[ \frac{g_q (0)\overline{g}_q (0)}{g_q (\frac{\pi}{2N})
\overline{g}_q (\frac{\pi}{2N})} \right]^{M} 
\prod_{k=1}^{L}\frac{\sin(\frac{\pi}{2N}-v_{k})}{\sin v_{k}} 
\eeq

We shall use the set of functional equations for the eigenvalues 
of transfer matrices of Chiral Potts derived by Baxter, Bazhanov and Perk
\cite{pBBP,BS}. This functional relation appears in reference \cite{pBBP} as
Eq. (4.40) and has the following form,
\begin{equation}\label{e18}
\tilde{T}_{\bar{q}}=\sum_{m=0}^{N-1}c_{m,q}T_{U^{m}q}^{-1}T_{q}T_{U^{m+1}q}^{-1}X^{-m-1}
\end{equation}
where $\tilde{T}=TS$, $\bar{q}=(a_{\bar{q}},b_{\bar{q}})=UR^{-1}(a_{q},b_{q})$, and
\bd
c_{m,q}= \left( \left(\prod_{j=0}^{m-1}\frac{b_{p}-\omega^{j+1}a_{q}}{a_{p}-\omega^{j}a_{q}} \right) \cdot \left( \prod_{j=m+1}^{N-1}\frac{\omega (a_{p}-\omega^{j}a_{q})}{b_{p}-\omega^{j+1}a_{q}} \right) \cdot \left( \frac{N(b_{q}-b_{p})(b_{p}-a_{q})}{a_{p}b_{p}-\omega^{m}a_{q}b_{q}} \right) \right)^{M}
\ed
\beqar
\tilde{T}_{\bar{q}} &=& \sum_{s=0}^{\frac{N}{2}-1} \left( c_{2s,q} T_{U^{2s}q}^{-1}T_{q}T_{U^{2s+1}q}^{-1}X^{-2s-1} \right) + 
\sum_{s=0}^{\frac{N}{2}-1} \left( c_{2s+1,q} T_{U^{2s+1}q}^{-1}T_{q}T_{U^{2s+2}q}^{-1}X^{-2s-2} \right)  \nonumber \\
&=& \sum_{s=0}^{\frac{N}{2}-1} \left( \frac{c_{2s,q}}{A_{s,q}A_{s,q'}} \cdot T_{R^{2s}q}^{-1}T_{q}T_{R^{2s}(Uq)}^{-1} X^{-1} +
\frac{c_{2s+1,q}}{A_{s,q'}A_{s+1,q}} \cdot T_{R^{2s}(Uq)}^{-1} T_{q} T_{R^{2s+2}q}^{-1} X^{-1} \right) \nonumber
\eeqar
Define
\bd
p_{2s} = \frac{c_{2s,q}}{A_{s,q}A_{s,q'}} , \hskip 1.5cm  
d_{2s+1} = \frac{c_{2s+1,q}}{A_{s,q'}A_{s+1,q}} 
\ed

The independent inverse factors of $T_{R^{2s}q}$ and $T_{R^{2s}(Uq)}$ are
considered, and both sides of the above equation are multiplied by the appropriate
common factor so as to get rid of inverses of transfer matrix. The appropriate
factor is:
\beq
X \prod_{j=1}^{\frac{N}{2}} T_{R^{2j}q} \prod_{j=0}^{\frac{N}{2}-1} T_{R^{2j}(Uq)} 
\eeq
\newline

\noindent
After multiplying we get
\beqar
X \cdot \tilde{T}_{\bar{q}} \cdot \prod_{j=1}^{\frac{N}{2}} T_{R^{2j}q} T_{R^{2(j-1)}(Uq)} = \sum_{s=0}^{\frac{N}{2}-1} \left( p_{2s} \prod_{\stackrel{j=0}{j\neq s}}^{\frac{N}{2}} T_{R^{2j}q} \cdot \prod_{\stackrel{j=0}{j\neq s}}^{\frac{N}{2}-1} T_{R^{2j}(Uq)} + \right. && \nonumber \\
\left. d_{2s+1} \prod_{\stackrel{j=0}{j\neq (s+1)}}^{\frac{N}{2}} T_{R^{2j}q} \cdot \prod_{\stackrel{j=0}{j\neq s}}^{\frac{N}{2}-1} T_{R^{2j}(Uq)} \right) &&
\eeqar
\newline

\noindent
If one expresses $T_{p,q}$ and $T_{p,Uq}$ in terms of a complex parameter 
$u$, where
$u = q-p $ as $ k \rightarrow 0$,
\beqar
& T_q \rightarrow T_q \hfill & \hskip 2cm T_{Uq} \rightarrow T_{Uq}(u) \nonumber \\
& T_{R^{2k}q} \rightarrow T_{q}(u+\frac{k \pi}{N}) \hfill & \hskip 2cm T_{R^{2k}(Uq)} \rightarrow T_{(Uq)}(u+\frac{k \pi}{N}) 
\eeqar
with this parametrization we get 

\beqar
X \cdot \tilde{T}_{\bar{q}} \cdot \prod_{j=1}^{\frac{N}{2}} T_q(u+\frac{\pi j}{N}) \cdot \prod_{j=0}^{\frac{N}{2}-1} T_{Uq}(u+\frac{\pi j}{N})=
\sum_{s=0}^{\frac{N}{2}-1} \left( p_{2s} \prod_{\stackrel{j=0}{j\neq s}}^{\frac{N}{2}} T_q(u+\frac{\pi j}{N}) \prod_{\stackrel{j=0}{j\neq s}}^{\frac{N}{2}-1} T_{Uq}(u+\frac{\pi j}{N})\right. \nonumber \\
+ \left. d_{2s+1} \prod_{\stackrel{j=0}{j \neq s+1}}^{\frac{N}{2}} T_q(u+\frac{\pi j}{N}) \prod_{\stackrel{j=0}{j \neq s}}^{\frac{N}{2}-1} T_{Uq}(u+\frac{\pi j}{N}) \right) 
\eeqa

Let $v$ be a zero of $T_q$, i.e., $T_q(v) = 0$, then whenever 
$u = v - \pi k/N$, $T_q(u+\pi k/N) = 0$. 
Thus for $u = v - \pi k/N\;\;\; k \in \{1,2,\dots , \frac{N}{2}-1\}$ 
all but two terms vanish.
\beq
p_{2k} \prod_{\stackrel{j=0}{j\neq k}}^{\frac{N}{2}} T_q(u+\frac{\pi j}{N}) \prod_{\stackrel{j=0}{j\neq k}}^{\frac{N}{2}-1} T_{Uq}(u+\frac{\pi j}{N}) + d_{2k-1} \prod_{\stackrel{j=0}{j \neq k}}^{\frac{N}{2}} T_q(u+\frac{\pi j}{N}) \prod_{\stackrel{j=0}{j \neq k-1}}^{\frac{N}{2}-1} T_{Uq}(u+\frac{\pi j}{N}) = 0
\eeq
Cancelling the common factors we get
\beq
p_{2k}(u) \cdot T_{Uq}(u+\frac{\pi k}{N} - \frac{\pi}{N}) + d_{2k-1}(u) \cdot T_{Uq}(u+\frac{\pi k}{N}) = 0 \nonumber
\eeq
whence
\beq
\frac{T_{Uq}(v)}{T_{Uq}(v-\frac{\pi}{N})} = - \frac{p_{2k}(v - \frac{\pi k}{N})}{d_{2k-1}(v - \frac{\pi k}{N})}
\eeq
Recalling the expression for $T_{Uq}$,
\beqar
\prod_{j=1}^{L_{Uq}} \frac{\sin(v_i - \bar{v}_j)}{\sin(v_i - \bar{v}_j - \frac{\pi}{N})} \;=\; - \left( \frac{g_{_{Uq}}(v_i) \cdot \bar{g}_{_{Uq}}(v_i)}{g_{_{Uq}}(v_i - \frac{\pi}{N}) \cdot \bar{g}_{_{Uq}}(v_i - \frac{\pi}{N})} \right)^M \cdot \frac{p_{2k}(v_i - \frac{\pi k}{N})}{d_{2k-1}(v_i - \frac{\pi k}{N})} \nonumber \\
\eeqar
The ratio of $g_{_{Uq}}$'s can be obtained as
\beqa
&& g_{_{Uq}}(v_i) = \prod_{j=1}^{\frac{N}{2}-1} \sin( \frac{\pi j}{N} + \frac{\pi}{2N} + v_i), \hskip 1cm
g_{_{Uq}}(v_i - \frac{\pi}{N}) = \prod_{j=0}^{\frac{N}{2}-2} \sin( \frac{\pi j}{N} + \frac{\pi}{2N} + v_i) \nonumber \\
&& \frac{g_{_{Uq}}(v_i)}{g_{_{Uq}}(v_i - \frac{\pi}{N})} = \frac{\sin( \frac{\pi}{2} - \frac{\pi}{N} + \frac{\pi}{2N} + v_i)}{\sin( 0 + \frac{\pi}{2N} + v_i)} 
=  \frac{\cos(v_i - \frac{\pi}{2N})}{\sin(v_i + \frac{\pi}{2N})}
\eeqa
Similarly the ratios of $\bar{g}_{_{Uq}}$ is found as,
\beqa
&& \overline{g}_{_{Uq}}(v_i) = \prod_{j=1}^{\frac{N}{2}-1} \sin( \frac{\pi j}{N} - v_i), \hskip 1cm
\overline{g}_{_{Uq}}(v_i - \frac{\pi}{N}) = \prod_{j=2}^{\frac{N}{2}}\sin( \frac{\pi j}{N} - v_i) \nonumber\\
&& \frac{\overline{g}_{_{Uq}}(v_i)}{\overline{g}_{_{Uq}}(v_i - \frac{\pi}{N})} = \frac{\sin( \frac{\pi}{N} - v_i)}{\sin( \frac{\pi}{2} -v_i)} 
= (-1) \cdot \frac{\sin(v_i + \frac{\pi}{N})}{\cos(v_i)}
\eeqa
Using these results for the ratios of $g_{Uq}$ and those of $(p_{2k}/d_{2k-1})$
we finally get the Bethe equations \cite{pbethe}
\beq
\prod_{j=1}^{L_{Uq}} \frac{\sin(v_i - \bar{v}_j)}{\sin(v_i - \bar{v}_j - \frac{\pi}{N})} = (-1)^{M+1} \left[ \frac{\sin 2(v_i - \frac{\pi}{2N})}{\sin(2 v_i)} \right]^{2M}
\eeq
\newline

\noindent
Let $\bar{v}$ be a zero of $T_{_{Uq}}$, i.e., $T_{_{Uq}}(\bar{v}) = 0$, 
then whenever $u = \bar{v} - \pi k/N$, $T_q(u+\pi k/N) = 0$. For 
$u = \bar{v} - \pi k/N\;\;\; k \in \{1,2,\dots , \frac{N}{2}-1\}$ all 
but two terms vanish
\beq
p_{2k} \prod_{\stackrel{j=0}{j\neq k}}^{\frac{N}{2}} T_q(u+\frac{\pi j}{N}) \prod_{\stackrel{j=0}{j\neq k}}^{\frac{N}{2}-1} T_{Uq}(u+\frac{\pi j}{N}) + d_{2k+1} \prod_{\stackrel{j=0}{j \neq k+1}}^{\frac{N}{2}} T_q(u+\frac{\pi j}{N}) \prod_{\stackrel{j=0}{j \neq k}}^{\frac{N}{2}-1} T_{Uq}(u+\frac{\pi j}{N}) = 0
\eeq
Cancelling the common factors we get
\beqar
&& p_{2k}(u) \cdot T_{q}(u+\frac{\pi k}{N} + \frac{\pi}{N}) + d_{2k+1}(u) \cdot T_{q}(u+\frac{\pi k}{N}) = 0 \nonumber  \\
&& \frac{T_{q}(\bar{v})}{T_{q}(\bar{v}+\frac{\pi}{N})} = - \frac{p_{2k}(\bar{v} - \frac{\pi k}{N})}{d_{2k+1}(\bar{v} - \frac{\pi k}{N})}
\eeqar
The ratios of $g_{_{q}}$'s can be obtained as
\beqar
&& g_{_{q}}(\bar{v}_i) = \prod_{j=1}^{\frac{N}{2}} \sin( \frac{\pi j}{N} - \frac{\pi}{2N} + \bar{v}_i), \hskip 1cm 
g_{_{q}}(\bar{v}_i + \frac{\pi}{N}) = \prod_{j=2}^{\frac{N}{2}+1} \sin( \frac{\pi j}{N} - \frac{\pi}{2N} + \bar{v}_i) \nonumber \\
&& \frac{g_{_{q}}(\bar{v}_i)}{g_{_{q}}(\bar{v}_i + \frac{\pi}{N})} = \frac{\sin( \frac{\pi}{N} - \frac{\pi}{2N} + \bar{v}_i)}{\sin( \frac{\pi}{N} + \frac{\pi}{N} + \bar{v}_i)} 
=  \frac{\sin(\bar{v}_i + \frac{\pi}{2N})}{\cos(\bar{v}_i + \frac{\pi}{2N})}
\eeqar
Similarly the ratios of $\overline{g}_{_{q}}$'s is found as,
\beqa
&& \overline{g}_{_{q}}(\bar{v}_i) = \prod_{j=1}^{\frac{N}{2}} \sin( \frac{\pi j}{N} - \bar{v}_i), \hskip 1cm
\overline{g}_{_{q}}(\bar{v}_i + \frac{\pi}{N}) = \prod_{j=0}^{\frac{N}{2}-1} \sin( \frac{\pi j}{N} - \bar{v}_i) \nonumber \\
&& \frac{\overline{g}_{_{q}}(\bar{v}_i)}{\overline{g}_{_{q}}(\bar{v}_i + \frac{\pi}{N})} = \frac{\sin( \frac{\pi}{2} - \bar{v}_i)}{\sin(0 -v_i)}
= (-1) \cdot \frac{\cos(\bar{v}_i)}{\sin(\bar{v}_i)}
\eeqa
Using these results for the ratios of $g_{q}$ and those of $p_{2k}/d_{2k+1}$
we finally get
\bd
\prod_{j=1}^{L_{q}} \frac{\sin(\bar{v}_i - v_j)}{\sin(\bar{v}_i - v_j + \frac{\pi}{N})} = (-1)^{M+1} \\
\ed

In order to cast the BAE's for even case in a simpler
(and standard) form, we make a change of variables.
\beq
v_j = i \lambda_j + \frac{\pi}{4N}, \hskip 1.5cm
\bar{v}_j = i \bar{\lambda}_j - \frac{\pi}{4N} 
\eeq
The BAE's in terms of these new variables are 
\beqar
\prod_{k=1}^{L_{Uq}} \frac{\sinh (\lambda_j -\bar{\lambda}_k - i \gamma)}{\sinh (\lambda_j -\bar{\lambda}_k + i \gamma)} &=& (-1)^{M+1} \left[ \frac{\sinh 2(\lambda_j + i s \gamma)}{\sinh 2(\lambda_j - i s \gamma)} \right]^{2M} \\
\prod_{k=1}^{L_{q}} \frac{\sinh (\bar{\lambda}_j -\lambda_k - i \gamma)}{\sinh (\bar{\lambda}_j -\lambda_k + i \gamma)} &=& (-1)^{M+1} 
\eeqar
where $\gamma=\frac{\pi}{2N}$  and  $s=\frac{1}{2}$.
From the numerical study for even spin BAE-s, it was found that 
$\lambda_j s$ are related to one another. In fact
\beq
\forall \;\;\; \lambda_j \;\;\; \exists \;\;\;\; \lambda_j + i \pi/2 \;\; \mbox{mod}(\pi)
\eeq
This allows us to group $\lambda_j$ such that $\lambda_j \in [-\pi/4, \pi/4]$. 
Using transformation rules for the hyperbolic functions one can rewrite
the expressions in terms of a new variable $\chi_j = 2 \lambda_j$.
The LHS of BAE(1) becomes
\beqar
&&\prod_{k=1}^{\frac{L_{Uq}}{2}} \frac{\sinh (\lambda_j -\bar{\lambda}_k-i \gamma)}{\sinh (\lambda_j -\bar{\lambda}_k+i \gamma)} \cdot \frac{\sinh (\lambda_j -\bar{\lambda}_k-i \gamma- \frac{i \pi}{2})}{\sinh (\lambda_j -\bar{\lambda}_k+i \gamma-\frac{i \pi}{2})}  \nonumber \\
&& \;\;\;\; = \prod_{k=1}^{\frac{L_{Uq}}{2}} \frac{\sinh 2(\lambda_j -\bar{\lambda}_k-i \gamma)}{\sinh 2(\lambda_j -\bar{\lambda}_k+i \gamma)} = \prod_{k=1}^{\frac{L_{Uq}}{2}} \frac{\sinh (\chi_j -\bar{\chi}_k-2 i \gamma)}{\sinh (\chi_j -\bar{\chi}_k+2 i \gamma)} 
\eeqar
RHS of BAE(1) is rewritten in terms of variables $\chi_j$
\beq
\left( -1 \right)^{M+1} \left( \frac{\sinh (\chi_j+ 2 i s \gamma)}{\sinh (\chi_j- 2 i s \gamma)} \right)^{2 M}
= \left( -1 \right)^{M+1} \left( \frac{\sinh (\chi_j+  i \gamma)}{\sinh (\chi_j- i \gamma)} \right)^{2 M},  \;\;\; \mbox{since} \;\; s=\frac{1}{2}
\eeq
A similar transformation is done for BAE(2). Hence the BAE equations become
\beqar
\prod_{k=1}^{\frac{L_{Uq}}{2}} \frac{\sinh (\chi_j -\bar{\chi}_k-2 i \gamma)}{\sinh (\chi_j -\bar{\chi}_k+2 i \gamma)} &=& \left( -1 \right)^{M+1} \left( \frac{\sinh (\chi_j+  i \gamma)}{\sinh (\chi_j- i \gamma)} \right)^{2 M}  \\
\prod_{k=1}^{\frac{L_{q}}{2}} \frac{\sinh (\bar{\chi}_j -\chi_k-2 i \gamma)}{\sinh (\bar{\chi}_j -\chi_k+2 i \gamma)} &=& \left( -1 \right)^{M+1} 
\eeqar
\newline

The eigenvalues $\Lambda_Q (u)$ of the transfer matrix are given in terms of
these zeroes $\chi_j$s as \cite{Sray01},
\beq
\Lambda_{Q}(u) = \left[ \frac{g_q (0) \overline{g}_q (0)}{g_q (u) \overline{g}_q (u)} \right]^{M}
\prod_{k=1}^{\frac{L}{2}} \frac{\sin(2 u-i \chi_{k} - \frac{\pi}{2N})}{\sin (i \chi_{k} + \frac{\pi}{2N})} 
\eeq

where,
\beq
g_q (u) = \prod_{j=1}^{N/2}\sin(\frac{\pi j}{N}-\frac{\pi}{2N}+u), \hskip 1cm
\overline{g}_q (u) = \prod_{j=1}^{N/2}\sin(\frac{\pi j}{N}-u) 
\eeq

%%%%%%%%%%%%%%%%%%%%%%%%%%%%%%%%%%%%%%%%%%%
% 	Now begins the The numerical Finite Size Computation
%%%%%%%%%%%%%%%%%%%%%%%%%%%%%%%%%%%%%%%%%%%

\section{Study of finite size systems}

The BAE in the present model differs from the standard form.
The sign in front of $\gamma$ on the right hand side (RHS)
of  BAE (1) is reversed.
That is to say that the RHS is equal to the inverse of what usually 
is known to
be the RHS. The second of the coupled pair, BAE (2) is even more striking. 
Though the LHS is still
coupled, the RHS is independent of the spectral variable ! Understanding 
the real significance 
of these peculiarities can help enormously in solving the problem.  \\

The transfer matrix for the FZM is constructed from the FZM Boltzmann weights (BW) as:
\beq
T_q^{\mbox{\boldmath $n,n'$}}(u) = T_{p,q}^{\mbox{\boldmath $n,n'$}}(u) = \prod_{k=1}^{M}\overline{W}_{pq} (n_{k}-n_{k}'|u) W_{pq} (n_{k}-n_{k+1}'|u).
\eeq
where $M$ is the number of sites in each row and periodic boundary condition
is implied. These transfer matrices for different spectral variable $u$ form a
commuting family. Transfer matrix acts on vectors defined in terms of 
spin indices along a row (or diagonal) \cite{Sray01} or the spin configuration
$\;\;{\mbox{\boldmath $n$}} =|n_{1},n_{2},\dots n_{M}>$. \\

There is an associated transfer matrix $T_{p,Uq}$ which corresponds to 
a conjugate set of Boltzmann weights \cite{srjmp},
\beq
T_{Uq}^{\mbox{\boldmath $n,n'$}}(u) = T_{p,Uq}^{\mbox{\boldmath $n,n'$}}(u) = \prod_{k=1}^{M}\overline{W}_{pUq} (n_{k}-n_{k}'|u) W_{pUq} (n_{k}-n_{k+1}'|u).
\eeq
The shift or translation operator $\widehat{S}$ is defined by its 
action on a state function or spin configuration 
$\;\;{\mbox{\boldmath $n$}} =|n_{1},n_{2},\dots n_{M}>$.
\beq
\widehat{S} |n_{1}, n_{2}, n_{3}, \dots n_{M}>  = |n_{M}, n_{1}, n_{2}, \dots n_{M-1} >
\eeq
Momentum $P$ is defined in terms of the shift operator as
\bd
e^{i P} = \widehat{S}^{-1}
\ed

The z-component of spin operator $\widehat{Z}_k$ and spin raising
operator $\widehat{X}_k$ corresponding to a given lattice site ($k$)
are defined by their 
action on a state function or spin configuration as
${\mbox{\boldmath $n$}} =|n_{1},n_{2},\dots n_{M}>$, $n_{k}=0,1 \dots N-1$.
\begin{eqnarray*}
\widehat{X}_{k}|n_{1} \dots n_{k} \dots n_{M}> & = &  |n_{1} \dots n_{k}+1 \dots n_{M}>\ \ \mbox{mod}(N)\\
\widehat{Z}_{k}|n_{1} \dots n_{k} \dots n_{M}> & = & \omega^{n_{k}}|n_{1}\dots n_{k}\dots n_{M}> \\
\end{eqnarray*}

The global spin raising operator is given by
\bd
\widehat{X}=\prod_{k=1}^{M} \widehat{X}_{k}
\ed

The spin $Q$ is defined in terms of $\widehat{X}$ as
\bd
e^{i Q} = \widehat{X}^{-1}
\ed

$T_{p,q}(u)$  and $T_{p,Uq}(u)$ commute with $\widehat{S}$, $\widehat{Z}_k$ 
and $\widehat{X}$.
Hence it is possible to make a spin and momentum sectorwise study 
of the problem.

The roots of the BAE  are studied by computing the eigenvalues
of the transfer matrices as meromorphic functions of $x = e^{\lambda}$. Since the
transfer matrices with different spectral parameter $\lambda$ commute, their 
eigenvectors are independent of the spectral parameter. Hence by taking a 
specific value of the spectral parameter one can determine the eigenvectors
numerically by diagonalizing the finite size transfer matrix. From the definition
of the eigenvalue equation one can express the eigenvalues as meromorphic functions
of $x$, as the entries of transfer matrix are polynomials in $x$ and the
eigenvectors are vectors with numerical (independent of $x$) entries. \\

However the problem being coupled, one needs to simultaneously diagonalize $T_q$ and $T_{Uq}$ or in 
other words find the eigenvalues corresponding to simultaneous eigenvectors of $T_q$ and $T_{Uq}$.
This fact by itself introduces a significantly higher level of difficulty over other numerical 
simulation of similar type (non coupled eqns.) e.g. Chiral Potts \cite{ADM}.
Coupled BAE and any simultaneous eigenvalue problem (eigenvalues corresponding to simultaneous 
eigenvectors) results in the same generic problem in computer algorithm. 
One has to develop efficient optimized codes for tackling this. \\

A detailed numerical study for chains of length $ M \le 8 $ \cite{Sray01} was 
done. The main observations of this numerical study are as follows:
\begin{itemize}
\item 1-String with both parities: $(1,v)\;\;$, $ v= \;(\pm 1) $
\item even length strings with positive parities: $ \;(n,+),\;\; n=2,4...,N $
\item non-string solutions      $\;\;\; Im (\lambda) \sim \pm \pi /3 $
\end{itemize}

The Ferromagnetic ground state is
a filled band of 2-strings, and the excitations consist of $(1+)$, $(1-)$
etc. The Antiferromagnetic ground state on the other hand is a filled band of
(1, $\pm$) and excitations are 2-strings with positive parity. \\

\begin{figure}[h]
\centerline{\psfig{figure=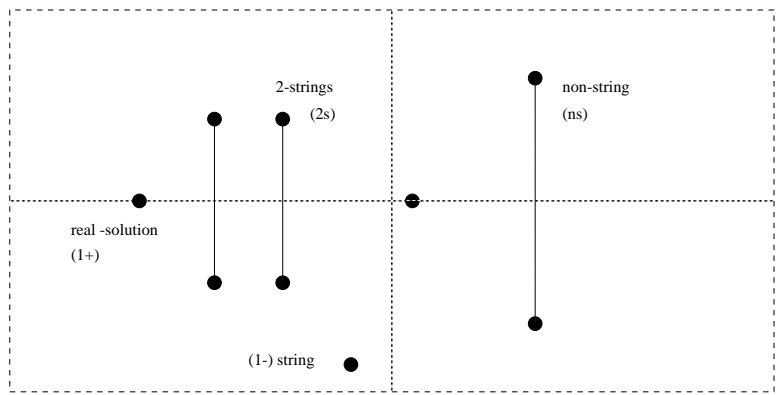,height=2in}}
\vspace{0.in}
\caption{String and non-string solutions for BAE}
\label{string.ps}
\end{figure}

It is remarkable that a good deal of insight into the nature of string solutions can be
obtained from finite systems of rather small size. The 2-strings are easily identified 
as having imaginary parts close to $\pi /4 $. One can also identify the real roots
and roots with negative parity $ Im (\lambda) = \pi /2 $. 
There are roots whose imaginary parts are not well approximated by $\pi /2$ or $\pi /4$. 
These roots do not seem to systematically approach the 2-string value as $M \rightarrow \infty $.
These roots are classified as non-strings. For space limitation $M=2$ and $M=3$ figures 
are presented in this paper. The imaginary parts of the non-string (ns) roots are  

\begin{tabbing}
MMMMM \= PPPPPPPPPPPP \= RRRRRRRRRRRRRRRRRRR  \kill
M=2 \> P=0 \> $-0.32198235 \pi i$  \\
\> \> $\;\;\;0.32198235 \pi i $  \\
M=3 \> P=0 \> $-0.29730902 \pi i $  \\
\> \> $\;\;\;0.29730902 \pi i $  \\
\> P=1 \> $\;\;\;0.179912 - 0.330545 \pi i $ \\
\> \> $\;\;\;0.179912 + 0.33054515 \pi i $ \\
\> P=2 \> $-0.179912 - 0.330545 \pi i $ \\
\> \> $-0.179912 + 0.33054515 \pi i $ \\
M=4 \> P=0 \> $-0.33780988 \pi i $ \\
\> \> $\;\;\;0.33780988 \pi i $ \\
\> \> $-0.348355 - 0.342016 \pi i $ \\
\> \> $-0.348355 + 0.34201621 \pi i $ \\
\> \> $\;\;\;0.348355 - 0.342016 \pi i $ \\
\> \> $\;\;\;0.348355 + 0.34201621 \pi i $ \\
\> \> $-0.28324874 \pi i $  \\
\> \> $\;\;\;0.28324874 \pi i $ \\
\> P=1 \> $-0.258619 - 0.38794057 \pi i $ \\
\> \> $-0.258619 + 0.38794057 \pi i $ 
\end{tabbing}

The following tables show the classification of roots for spin-4, for lattice sizes
2, 3 and 4 in the $Q=0$ sector. The first column shows the momentum $P$. The second column
gives the roots of the BAE in the variable $\chi$. The content or the type of root is
identified in column three. Column four gives the corresponding integer that appears 
in the BAE. In column five the energy calculated for the corresponding eigenvalue
for Transfer Matrix is given. \\

{\bf Table 1}: Classification of roots and integers for $M=2$  \\

\begin{tabular}{|c|l|c|c|r|}  \hline 
P &  $\lambda_k = \ln x_k$ & Content & $I_k$ & Energy \\ \hline
& $-0.216337 - 0.227395 \pi i $ & (2s) & $0.5$ & \\
& $-0.216337 + 0.22739469 \pi i $ &    &   & \\
& $0.216337 - 0.227395 \pi i $ & (2s) & $-0.5$ & \\ 
0& $0.216337 + 0.22739469 \pi i $ &   &   & $-6.24622$  \\ \hline
& $-0.265319$ & (1+) & $-0.5$ & \\
& $-0.32198235 \pi i $ & (ns) & $0$ & \\
& $0.32198235 \pi i $ &    &   & \\
0& $0.265319$ & (1+) & $0.5$ & $4.$ \\ \hline
& $-0.440687$ & (1+) & $-1.$ & \\
& $0$ & (1+) & $0$ & \\
& $-0.5 \pi i $ & $(1-)$ & $0$ & \\
1& $0.440687$ & (1+) & $1.$ & $8.$ \\ \hline
\end{tabular}

\vskip .8cm

%%%%%%%%%%%%%%%%%%%%%%%%%%%%%%%%%%%%%%%%%%%
% 	Now begins the M3 table
%%%%%%%%%%%%%%%%%%%%%%%%%%%%%%%%%%%%%%%%%%%

{\bf Table 2}: Classification of roots and integers for $M=3$ \\

\begin{tabular}{|c|l|c|c|r|}  \hline 
P &  $\lambda_k = \ln x_k$ & Content & $I_k$ & Energy \\ \hline
& $-0.325636 - 0.228897 \pi i $ & (2s) & $1.$ & \\
& $-0.325636 + 0.22889664 \pi i $ &    &   & \\
& $-0.23986244 \pi i $ & (2s) & $0$ & \\
& $0.23986244 \pi i $ &    &   & \\
& $0.325636 - 0.228897 \pi i $ & (2s) & $-1.$ & \\ 
0& $0.325636 + 0.22889664 \pi i $ &   &   & $-8.36945$  \\ \hline
& $-0.388991$ & (1+) & $-1.$ & \\
& $-0.267718 \pi i $ & (2s) & $0$ & \\
& $0.267718 \pi i $ &    &   & \\
& $-0.29730902 \pi i $ & (ns) & $0$ & \\
& $0.29730902 \pi i $ &    &   & \\
0& $0.388991$ & (1+) & $1.$ & $-1.65686$ \\ \hline
\end{tabular}

\newpage

{\bf Table 2 Contd.}: Classification of roots and integers for $M=3$ \\

\begin{tabular}{|c|l|c|c|r|}  \hline 
P &  $\lambda_k = \ln x_k$ & Content & $I_k$ & Energy \\ \hline
& $-0.570869$ & (1+) & $-1.5$ & \\
& $-0.213538 - 0.5 \pi i $ & $(1-)$ & $0$ & \\
& $-0.149008$ & (1+) & $-0.5$ & \\
& $0.0873277$ & (1+) & $0.5$ & \\
& $0.423044 - 0.242196 \pi i $ & (2s) & $-1.5$ & \\ 
0& $0.423044 + 0.24219619 \pi i $ &   &   & $8.$  \\ \hline
& $-0.768026$ & (1+) & $-2.$ & \\
& $-0.241758$ & (1+) & $-1.$ & \\
& $-0.24949865 \pi i $ & (2s) & $0$ & \\
& $0.24949865 \pi i $ &    &   & \\
& $0.241758$ & (1+) & $1.$ & \\
0& $0.768026$ & (1+) & $2.$ & $4.48683$ \\ \hline
& $-0.353241 - 0.236642 \pi i $ & (2s) & $1.$ & \\
& $-0.353241 + 0.23664231 \pi i $ &    &   & \\
& $-0.0132479$ & (1+) & $0$ & \\
& $0.179912 - 0.330545 \pi i $ & (ns) & $0$ & \\
& $0.179912 + 0.33054515 \pi i $ &    &   & \\
1& $0.359907$ & (1+) & $1.$ & $3.37155$ \\ \hline
& $-0.540092$ & (1+) & $-1.5$ & \\
& $-0.129032$ & (1+) & $-0.5$ & \\
& $-0.0725776 - 0.5 \pi i $ & $(1-)$ & $0$ & \\
& $0.117714 - 0.250431 \pi i $ & (2s) & $-0.5$ & \\
& $0.117714 + 0.25043057 \pi i $ &    &   & \\
1& $0.506275$ & (1+) & $1.5$ & $3.10102$ \\ \hline
& $-0.692847$ & (1+) & $-2.$ & \\
& $-0.209487 - 0.250419 \pi i $ & (2s) & $1.$ & \\
& $-0.209487 + 0.25041875 \pi i $ &    &   & \\
& $0.0217873$ & (1+) & $0$ & \\
& $0.2683$ & (1+) & $1.$ & \\
1& $0.821734$ & (1+) & $2.$ & $6.82843$ \\ \hline
& $-0.856212$ & (1+) & $-2.$ & \\
& $-0.287304$ & (1+) & $-1.$ & \\
& $-0.0386368$ & (1+) & $0$ & \\
& $0.185678$ & (1+) & $1.$ & \\
& $0.375937 - 0.5 \pi i $ & $(1-)$ & $-0.5$ & \\
1& $0.620538$ & (1+) & $2.$ & $12.899$ \\ \hline
& $-0.885356$ & (1+) & $-2.$ & \\
& $-0.304098$ & (1+) & $-1.$ & \\
& $-0.0535889$ & (1+) & $0$ & \\
& $0.16349$ & (1+) & $1.$ & \\
& $0.539776 - 0.247284 \pi i $ & (2s) & $-1.$ & \\ 
1& $0.539776 + 0.24728422 \pi i $ &   &   & $11.4569$  \\ \hline
& $-0.359907$ & (1+) & $-1.$ & \\
& $-0.179912 - 0.330545 \pi i $ & (ns) & $0$ & \\
& $-0.179912 + 0.33054515 \pi i $ &    &   & \\
& $0.0132479$ & (1+) & $0$ & \\
& $0.353241 - 0.236642 \pi i $ & (2s) & $-1.$ & \\ 
2& $0.353241 + 0.23664231 \pi i $ &   &   & $3.37155$  \\ \hline
\end{tabular}

\newpage

{\bf Table 2 Contd.}: Classification of roots and integers for $M=3$ \\

\begin{tabular}{|c|l|c|c|r|}  \hline 
P &  $\lambda_k = \ln x_k$ & Content & $I_k$ & Energy \\ \hline
& $-0.506275$ & (1+) & $-1.5$ & \\
& $-0.117714 - 0.250431 \pi i $ & (2s) & $0.5$ & \\
& $-0.117714 + 0.25043057 \pi i $ &    &   & \\
& $0.0725776 - 0.5 \pi i $ & $(1-)$ & $0$ & \\
& $0.129032$ & (1+) & $0.5$ & \\
2& $0.540092$ & (1+) & $1.5$ & $3.10102$ \\ \hline
& $-0.539776 - 0.247284 \pi i $ & (2s) & $1.$ & \\
& $-0.539776 + 0.24728422 \pi i $ &    &   & \\
& $-0.16349$ & (1+) & $-1.$ & \\
& $0.0535889$ & (1+) & $0$ & \\
& $0.304098$ & (1+) & $1.$ & \\
2& $0.885356$ & (1+) & $2.$ & $11.4569$ \\ \hline
& $-0.620538$ & (1+) & $-2.$ & \\
& $-0.375937 - 0.5 \pi i $ & $(1-)$ & $0.5$ & \\
& $-0.185678$ & (1+) & $-1.$ & \\
& $0.0386368$ & (1+) & $0$ & \\
& $0.287304$ & (1+) & $1.$ & \\
2& $0.856212$ & (1+) & $2.$ & $12.899$ \\ \hline
& $-0.821734$ & (1+) & $-2.$ & \\
& $-0.2683$ & (1+) & $-1.$ & \\
& $-0.0217873$ & (1+) & $0$ & \\
& $0.209487 - 0.250419 \pi i $ & (2s) & $-1.$ & \\
& $0.209487 + 0.25041875 \pi i $ &    &   & \\
2& $0.692847$ & (1+) & $2.$ & $6.82843$ \\ \hline
\end{tabular}

%%%%%%%%%%%%%%%%%%%%%%%%%%%%%%%%%%%%%%%%%%%
% 	Now begins the FM ground state
%%%%%%%%%%%%%%%%%%%%%%%%%%%%%%%%%%%%%%%%%%%

\section{Free Energy in the Ferromagnetic case for $N$ even}

From the numerical study one can identify that the Ferromagnetic (FM) ground
state corresponds to a filled band of $N/2$ strings of positive parity
for $T_q$ and a filled band of 1-string of negative parity for $T_{Uq}$. 
This vector always falls in the $P=0$ sector as is expected. A further study 
upto 6 sites reveals that this remains true. \\

The LHS of the first of Bethe Ansatz Equations, BAE(1), is given by:
\beq
\prod_{k=1}^{\frac{L_{_{Uq}}}{2}} \frac{\sinh(\chi_j-\bar{\chi}_k-2i \gamma)}{\sinh(\chi_j-\bar{\chi}_k+2i \gamma)} = (-1)^{M+1} \left[ \frac{\sinh(\chi_j+ \gamma i)}{\sinh(\chi_j- \gamma i)} \right]^{2 M}
\eeq

For the Ferromagnetic case we made the assumption that the 
ground state corresponds to $N/2$ strings with positive parity
for $T_q$ and 1-strings of negative parity for $T_{Uq}$.
\beqar
\chi_{\alpha,v}^{n,l} &=& \chi_{\alpha}^n + 2 \gamma (n+1-2 l) i \\
\bar{\chi}_{\alpha,v}^{n,l} &=& \bar{\chi}_{\alpha}^n - \frac{i \pi}{2} 
\eeqar
Define $x$ and $\bar{x}$ by $\chi = 2 \gamma x$ and  $\bar{\chi} = 2 \gamma \bar{x}$.
If $M^{(n)}$ denotes the number of n-strings, we
get for the left hand side (LHS) of the BAE(1) \\
\bd
\prod_{l=1}^{n} \frac{\sinh 2 \gamma (x_{\alpha}^n-\bar{x}_{\beta}+(n+1-2l)i-i-p_0 i)}{\sinh 2 \gamma (x_{\alpha}^n-\bar{x}_{\beta}+(n+1-2l)i+i-p_0 i)}
 = \frac{\sinh 2 \gamma (x_{\alpha}^n-\bar{x}_{\beta}-n i-p_0 i)}{\sinh 2 \gamma (x_{\alpha}^n-\bar{x}_{\beta}+n i-p_0 i)}
\ed

Taking product over the string 
elements of $\chi_{\alpha}$
the first BAE becomes decoupled and is given in terms of the 
variables for $T_q$ alone.
\beq
(-1)^{(M+1)n} \left[ \prod_{l=1}^{n} \frac{\sinh 2 \gamma (x_{\alpha}^n+(n+1-2l+\frac{1}{2}) i)}{\sinh 2 \gamma (x_{\alpha}^n-(n+1-2l+\frac{1}{2}) i)} \right]^{2 M} = (-1)^{M+1}
\eeq

\noindent
After multiplying for the elements of a string, the 
second BAE becomes
\beq
\prod_{k=1}^{\frac{L_q}{2}} \prod_{l=1}^{n} \frac{\sinh 2 \gamma (\bar{x}_{\alpha}-x_{\beta}^n-(n+1-2l)i-i-p_0 i)}{\sinh 2 \gamma (\bar{x}_{\alpha}-x_{\beta}^n-(n+1-2l)i+i-p_0 i)} = (-1)^{(M+1)+\frac{n L_q}{2}} \nonumber \\
\eeq

Thus we have only one set of BAE which 
involves the zeros of the transfer matrix $T_q$.
\noindent
From the above equation taking natural logarithm of both sides we get
\bd
2 M \sum_{l=1}^{n} i \cdot \ln \left( \frac{\sinh(\chi_{\alpha} + 
2 \gamma (n+1-2l+\frac{1}{2}))}{\sinh(\chi_{\alpha} - a \gamma 
(n+1-2l+\frac{1}{2}))} \right) = \pi I_{\alpha} 
\ed

Defining the density of string centers for the zeros of $T_q$ by
\beq
\rho(\chi) = \lim_{M \rightarrow \infty} \frac{1}{M(\chi_{k+1}-\chi_k)}
\eeq
we get,
\beq
\rho(\chi)= \frac{1}{\pi} \Theta_{(\frac{N}{2},+)}^{(1)'}(\chi)
\eeq
where 
\bd
\Theta_{(\frac{N}{2},+)}^{(1)}(\chi) \doteq \sum_{l=1}^{n} 2 \phi(\chi,n+\frac{1}{2}-2l,+)
\ed
and prime on $\Theta_{(\frac{N}{2},+)}^{(1)}(\chi)$ denotes differentiation with respect to the variable $\chi$.
Here the function $\phi$, as defined by Takahashi and Suzuki \cite{TS} is,
\bd
\phi (x,n,v) \doteq i \cdot \ln(g(x,n,v)) \;\;\; ; \;\;\;\;\;\;\;\;\;\;\;\;\; g(x,n,v) \doteq \frac{\sinh 2 \gamma (x + n i + p_0 i)}{\sinh 2 \gamma (x - n i + p_0 i)}
\ed

Evaluating the sum over $l$ in the Fourier space,
\beqar
\tilde{\Theta}_{(\frac{N}{2},+)}^{(1)'}(k) &=& 2 \pi \frac{\sinh( \frac{\pi k}{2} - \frac{\pi k}{2 N})}{\sinh(\frac{\pi k}{N})}  \\
\tilde{\rho}(k) &=& 4 \frac{\sinh( \frac{\pi k}{2} - \frac{\pi k}{2 N})}{\sinh(\frac{\pi k}{N})}
\eeqar

By inverse Fourier transform we get 
\beqa
\rho(\chi) &=& \frac{1}{2 \pi} \int_{- \infty}^{ + \infty} dk e^{-i k \chi} \tilde{\rho}(k) \nonumber 
\eeqa

The free energy for the Ferromagnetic ground state is defined as

\bd
f_0 (u)  \doteq  \lim_{M \rightarrow \infty} \left( - \frac{1}{M} \ln \Lambda_0 (u) \right) = \prod_{\alpha=1}^{M^{\frac{N}{2}}} \prod_{l=1}^{\frac{N}{2}} \frac{ \sin ( 2u- i \chi_{\alpha} + x(l) )}{ \sin ( i \chi_{\alpha} - x(l) )} \nonumber 
\ed
Replacing the summation by an integral over the symmetrically placed
string centers,
\beq
f_0 (u) = - \frac{1}{2} \int_{- \infty}^{\infty} d\chi \rho (\chi) \sum_{l=1}^{\frac{N}{2}} \ln \left( \frac{\cosh (2 \chi) -\cos (4 u + 2 x(l))}{\cosh (2 \chi) -\cos (2 x(l))} \right)
\eeq

Transforming to the Fourier space and utilizing the expression and 
properties of $\tilde{\rho}(k)$ we get 
\beq
f_0 (u) = \int_{- \infty}^{\infty} \frac{4 \;\; dk}{k} \frac{ \sinh(k \pi - \frac{k \pi}{N}) \sinh(2 k u) \sinh( 2 k u - k \pi - \frac{k \pi}{N})}{\sinh ^2 (\frac{2 k \pi}{N})}
\eeq

%%%%%%%%%%%%%%%%%%%%%%%%%%%%%%%%%%%%%%%%%%%
% 	Now begins the FM excitation
%%%%%%%%%%%%%%%%%%%%%%%%%%%%%%%%%%%%%%%%%%%

\section{Excitation on FM ground state}
We have seen in the previous section that the FM ground state is given by a filled band of $(2s)$
strings. Consider the $Z_{(2s)}(\chi)$ function. In the general
case it should look like
\beq
Z_{(2s)} (\chi) = \frac{1}{2 \pi} \Theta_{(2s)}^{(1)} (\chi) - \frac{1}{2 \pi M} \sum_{k} \sum_{\beta =1}^{M^{(k)}} \Theta_{(2s,k)}^{(2)} (\chi - \chi_{\beta}^k)
\eeq

The density of $(2s)$ vacancies is given by
\beq
\sigma_{(2s)} (\chi) \doteq - Z_{(2s)}'(\chi)
\eeq

The vacancy density and the density of (2s) particles is related by
\beq
\sigma_{(2s)} (\chi) = \rho_{(2s)} (\chi) + \frac{1}{M} \; \sum_{\beta=1}^{M^{(2s)}_h} \delta \left(\chi - \chi^{(2s)h}_{\beta} \right)
\eeq
where $\chi^{(2s)h}_{\beta}$ are the position of the holes.  \\
Thus
\beqar
- \sigma_{(2s)} (\chi) = \frac{1}{2 \pi} \Theta_{(2s)}^{(1)'}(\chi) - \frac{1}{2 \pi M} \sum_{k \neq (2s)} \sum_{\beta =1}^{M^{(k)}} \Theta_{(2s,k)}^{(2)'}  (\chi - \chi_{\beta}^k)-  \nonumber \\
\frac{1}{2 \pi} \int \Theta_{(2s,2s)}^{(2)'}(\chi - \mu) d\mu
+ \frac{1}{2 \pi M} \sum_{h=1}^{M^{(2s)}_h} \Theta_{(2s,2s)}^{(2)'}(\chi - \chi^{(2s)h}_{\beta})
\eeqar

The above equation can be interpreted as a collection of terms 
contributing to $(2s)$- ground state, $(2s)$- holes and excited
particles $ \sigma_{(2s)} = \sigma_{(2s)}^{(0)} +\sigma_{(2s)}^{(h)} + \sum_{j} \sigma_{(2s)}^{(j)}$ where $\sigma_{(2s)}^{(0)}$ is the same as $\rho_{(2s)}$ of the
last section.

The expressions for energy $(E)$ and momentum $(P)$ are
\beqar
E &=& \sum_{k=1}^{\frac{L}{2}} \cot \left(i \chi_k + \frac{\pi}{2 N}\right) - 2 M \sum_{k=1}^{\frac{N}{2}} \cot \left( \frac{\pi k}{N} \right) \\
e^{i P} &=& \prod_{k=1}^{\frac{L}{2}} \frac{\sinh \left( \chi_k + \frac{i \pi}{2 N}\right)}{\sinh \left( \chi_k - \frac{i \pi}{2 N}\right)} 
\eeqar

We obtain the energy of a state designated by a given set of
strings
\beqar
E &=& \sum_{\stackrel{k}{strings}} \sum_{\beta=1}^{M^{(k)}} \epsilon_k (\chi_{\beta}^{(k)}) \nonumber \\
&=& \int d \chi \sigma_{(2s)} (\chi) \epsilon_{(2s)} (\chi) - \sum_{\beta=1}^{M^{(2s)}_h} \epsilon_{(2s)} \left( \chi_{\beta}^{(2s)h} \right) + 
\sum_{\beta \neq k} \sum_{\beta=1}^{M^{(k)}} \epsilon \left( \chi_{\beta}^{(k)} \right)
\eeqar

The bare energies for n-string with parity $v$ is easily obtained
\beqar
\epsilon_{(n,v)} (\chi_{\alpha}) &=& \sum_{k=1}^{n} \cot \left( i \chi_{k,\alpha}^{(n,v)} + \frac{\pi}{2 N}\right) \nonumber \nonumber \\
&=& \sum_{k=1}^{n} \cot \left( i \chi_{\alpha}^{(n,v)} -2 \gamma (n+1-2k) - \frac{\pi}{4} (1-v) + \frac{\pi}{2 N} \right) \nonumber \\
&=& \sum_{k=1}^{n} \tan \left(\frac{\pi}{2} - i \chi_{\alpha}^{(n,v)} + 2 \gamma (n+1-2k) + \frac{\pi}{4} (1-v) - \frac{\pi}{2 N} \right) \nonumber
\eeqar

One can separate the real and imaginary parts of this expression. 
This helps in determining whether we require additional constraints
on rapidities $\chi_j$ to ensure reality of the total energy. \\

Numerical study showed that there exist several spurious
solutions and only a subset of them, corresponding to a specific
choice of counting numbers $I_{\alpha}^{(j)}$, is admissible.
Non-string solutions exist, however they are not
as numerous. From the numerical study  we make
the assumption that the elementary excitations over the FM (a
sea of $(2s)$-strings ) are (a) a pair of $(1+)$ strings , and (b) $(1+)$
and $(1-)$ strings. \\

The FM ground state is a filled band of $(2s)$ strings. The density
of ground state energy is
\beq
e_0 = \lim_{M \rightarrow \infty} \frac{E_0}{M} = \int d\chi \rho_{(2s)}(\chi) \epsilon_{(2s)}(\lambda) - 2 \sum_{k=1}^{\frac{N-1}{2}} \cot \left( \frac{\pi k}{N} \right) 
\eeq

The observed correlation between the integers suggest that the
rapidities corresponding to $(2s)$ and $(a)$ ought to be connected 
allowing cancellation of the imaginary part of the total energy.
\beqar
&& Im \left[ \epsilon_{(2s)}\right] = Im \left[ \epsilon_{(a)}\right]
\eeqar
It can be shown that, $Im \left[\epsilon_{(2s)} \right]= Im \left[ \epsilon_{(a)} \right] $
and $ Re \left[\epsilon_{(2s)}\right] = - Re \left[\epsilon_{(a)} \right] $.
\beq
Re \left[ \epsilon_{(a)} (\chi) \right] =  \frac{4}{\cosh (4 \chi)}
\eeq
Similar argument holds for (b)-type excitations, where 
$Im \left[\epsilon_{(2s)} \right] = Im \left[ \epsilon_{(b)} \right] $
and $ Re \left[\epsilon_{(2s)} \right] =  - Re \left[ \epsilon_{(b)} \right] $.
\beq
Re \left[ \epsilon_{(b)} (\chi) \right] =  \frac{4}{\cosh (4 \chi)} 
\eeq
One should note that the dressed energy and bare energy are
equal since the function coupling the ground state density
to excited states $\Theta_{(j,k)}^{(2)}$ is zero for $j=(2s)$.
Thus we arrive at
\beq
E = E_0 + \sum_{\beta = 1}^{M^{(a)}} 2 \epsilon_{(a)} \left( \chi_{\beta}^{(a)} \right) + \sum_{\beta = 1}^{M^{(b)}} 2 \epsilon_{(b)} \left( \chi_{\beta}^{(b)} \right)
\eeq
where $E_0$ is the ground state energy. \\

We now turn to the calculation of momentum. 
The momentum associated with a string of length $j$ and
parity $v$ is
found to be
\beqar
p_{(1+)} (\chi) = - \frac{1}{2} \Theta_{(1+)}^{(1)} (\chi)  \nonumber \\
p_{(j)} (\chi) = - \frac{1}{2} \Theta_{(j)}^{(1)} (\chi )  \;\;\;\;\; for \;\; j\neq (1+)
\eeqar
whence
\beq
p_{(a)} (\chi) = p_{(b)} (\chi) = 2 \arctan \left( \tanh \left( \frac{\chi}{2} \right) \right) + \pi
\eeq

These expressions are similar to the ones for 
non-string excitations in the Anti-Ferromagnetic case for odd spin FZM.
We get the dispersion relations for elementary excitations over the
Ferromagnetic ground state as,
\beqar
\epsilon_{(a)} ( p ) = 4  \sin \left( \frac{p}{2} \right) \nonumber \\
\epsilon_{(b)} ( p ) = 4  \sin \left( \frac{p}{2} \right) 
\eeqar

%%%%%%%%%%%%%%%%%%%%%%%%%%%%%%%%%%%%%%%%%%%
% 	Now begins the AFM ground state
%%%%%%%%%%%%%%%%%%%%%%%%%%%%%%%%%%%%%%%%%%%

\section{Free Energy in the Anti-Ferromagnetic case for $N$ even}
From the numerical study for finite lattices it was apparent that the 
Anti-Ferromagnetic (AFM) ground-state corresponds to a filled band of real
roots for both $T_q$ and $T_{Uq}$. In other words the AFM ground state
is a filled sea of $(1,+)$ strings for both families of transfer matrices. \\

The AFM ground state corresponds to real roots for both families of
transfer matrices. Hence we consider the natural logarithm of 
both sides of BAE-s. From BAE(1) we get,
\beq
\sum_{k=1}^{L_{Uq}/2} i \ln \left(\frac{\sinh(\chi_j-\bar{\chi}_k-2 i \gamma)}{\sinh(\chi_j-\bar{\chi}_k+2 i \gamma)}\right) = 2 M \cdot i \ln \left(\frac{\sinh(\chi_j+ i \gamma)}{\sinh(\chi_j- i \gamma)}\right) + 2 \pi I_{j} 
\eeq
with standard definition of $\rho_1(\chi)$ and $\rho_2(\chi)$,
\beqar
\rho_1 (\chi_{\alpha}) &=& \frac{1}{\pi} \Theta_1^{(1)'}(\chi_{\alpha}) - \frac{1}{2 M \pi} \sum_{\beta=1}^{\bar{M}} \Theta_1^{(2)'}(\chi_{\alpha}-\bar{\chi}_{\beta})  \nonumber \\
&=& \frac{1}{\pi} \Theta_1^{(1)'} (\chi_{\alpha}) - \frac{1}{2 M \pi} \int_{- \infty}^{ + \infty} d\bar{\mu} \Theta_1^{(2)'} (\chi-\bar{\mu}) \rho_1(\bar{\mu})
\eeqar
From BAE(2) we get,
\beq
\sum_{k=1}^{L_{q}/2} i \ln \left(\frac{\sinh(\bar{\chi}_j-\chi_k-2 i \gamma)}{\sinh(\bar{\chi}_j-\chi_k+2 i \gamma)}\right) = 2 \pi \bar{I_j} 
\eeq
In the continuum limit $M \rightarrow \infty$ we get,
\beq
\rho_2(\bar{\chi}) = \frac{1}{2 \pi} \int_{- \infty}^{ + \infty} d\mu \Theta_2^{(2)'}(\bar{\chi}_{\alpha}-\mu) \rho_1(\mu)
\eeq
\noindent
The above pair of BAE is solved as before by the Fourier transform method.
\beqar
&& \tilde{\rho}_1(k) = \frac{1}{\pi} \tilde{\Theta}_1^{(1)'}(k) - \frac{1}{2 \pi} \tilde{\Theta}_1^{(2)'}(k) \tilde{\rho}_2(k)   \\
&& \tilde{\rho}_2(k) = \frac{1}{2 \pi} \tilde{\Theta}_2^{(2)'}(k) \tilde{\rho}_1(k) 
\eeqar

\noindent
From the above two equations, 
\beq
\tilde{\rho}_1(k)=\frac{\sinh(\frac{\pi k}{2}) \cdot \sinh(\frac{\pi k}{2}-\frac{\pi k}{2N})}
{\sinh(\frac{\pi k}{N}) \cdot \sinh( \pi k+\frac{\pi k}{N})}
\eeq

\noindent
Following a similar procedure as shown in detail in the Ferromagnetic case, the
free energy for the Antiferromagnetic case is obtained as,
\beq
f= \frac{1}{2} \int_{- \infty}^{ + \infty} \frac{dk}{k} \frac{\tilde{\rho}_1(2k)}{\sinh(\pi k)}
\left[\cosh k(\pi-4u-\frac{\pi}{N})-\cosh k(\pi-\frac{\pi}{N})\right]
\eeq

\noindent
Substituting for $\tilde{\rho}_1(2k)$
\beqa
f= \frac{1}{2} \int_{- \infty}^{ + \infty} \frac{dk}{k} \frac{\sinh(\pi k-\frac{\pi k}{N})}
{\sinh(\frac{2 \pi k}{N}) \cdot \sinh(2 \pi k+\frac{2 \pi k}{N})} \cdot \cosh k(\pi-4u-\frac{\pi}{N}) - \nonumber \\
\int_{- \infty}^{ + \infty} \frac{dk}{k}\frac{\sinh(\pi k-\frac{\pi k}{N})}
{\sinh(\frac{2 \pi k}{N}) \cdot \sinh(2 \pi k+\frac{2 \pi k}{N})} \cdot \cosh k(\pi-\frac{\pi}{N})
\eeqa

%%%%%%%%%%%%%%%%%%%%%%%%%%%%%%%%%%%%%%%%%%%
% 	Now begins the AFM excitation
%%%%%%%%%%%%%%%%%%%%%%%%%%%%%%%%%%%%%%%%%%%

\section{Excitation on AFM ground state}

In the case of AFM, the ground state is a filled band of
$(1+)$ . The results of previous section suggest that excitations
appear as $(2s)$ strings when a hole is created in the $(1+)$ sea.
Consider the $Z_{(1+)}(\chi)$ function. In the general
case it should look like
\beq
Z_{(1+)} (\chi) = \frac{1}{2 \pi} \Theta_{(1+)}^{(1)} (\chi) - \frac{1}{2 \pi M} \sum_{k} \sum_{\beta =1}^{M^{(k)}} \Theta_{(1+,k)}^{(2)} (\chi - \chi_{\beta}^k)
\eeq

The density of $(1+)$ vacancies is given by
\beq
\sigma_{(1+)} (\chi) \doteq - Z_{(1+)}'(\chi)
\eeq

The vacancy density $ \sigma_{(1+)} (\chi)$ and the density of $(1+)$ particles $ \rho_{(1+)} (\chi)$ is related by
\beq
\sigma_{(1+)} (\chi) = \rho_{(1+)} (\chi) + \frac{1}{M} \; \sum_{\beta=1}^{M^{(1+)}_h} \delta \left(\chi - \chi^{(1+)h}_{\beta} \right)
\eeq
where $\chi^{(1+)h}_{\beta}$ are the position of the holes. \\
Thus
\beqar
- \sigma_{(1+)} (\chi) = \frac{1}{2 \pi} \Theta_{(1+)}^{(1)'}(\chi) - \frac{1}{2 \pi M} \sum_{k \neq (1+)} \sum_{\beta =1}^{M^{(k)}} \Theta_{(1+,k)}^{(2)'}  (\chi - \chi_{\beta}^k)-  \nonumber \\
\frac{1}{2 \pi} \int \Theta_{(1+,1+)}^{(2)'}(\chi - \mu) d\mu
+ \frac{1}{2 \pi M} \sum_{\beta=1}^{M^{(1+)}_h} \Theta_{(1+,1+)}^{(2)'}(\chi - \chi^{(1+)h}_{\beta})
\eeqar

The above equation can be interpreted as a collection of terms 
contributing to $(1+)$ ground state, $(1+)$ holes and excited
particles over the ground state sea of $(1+)$ string.
$ \sigma_{(1+)} = \sigma_{(1+)}^{(0)} +\sigma_{(1+)}^{(h)} + \sum_{j} \sigma_{(1+)}^{(j)}$, where $\sigma_{(1+)}^{(0)}$ is the same as $\rho_{(1+)}$ of the
last section. \\

\noindent
We obtain the energy of a state designated by a given set of
strings $\{k\}$, having $M^{(k)}$ strings of type ($k$) with
string centers at $\chi_{\beta}^{(k)}$,
\beqar
E &=& \sum_{\stackrel{k}{strings}} \sum_{\beta=1}^{M^{(k)}} \epsilon_k (\chi_{\beta}^{(k)}) \nonumber \\
&=& \int d \chi \sigma_{(1+)} (\chi) \epsilon_{(1+)} (\chi) - \sum_{\beta=1}^{M^{(1+)}_h} \epsilon_{(1+)} \left( \chi_{\beta}^{(1+)h} \right) + 
\sum_{k \neq (1+) } \sum_{\beta=1}^{M^{(k)}} \epsilon \left( \chi_{\beta}^{(k)} \right) \nonumber
\eeqar

From the numerical study we can make
the assumption that the elementary excitations over the AFM 
ground state, which is a sea of $(1+)$-strings, is given by
a set of $(2s)$ strings.  \\

The density of ground state energy for the AFM case is given by,
\beq
e_0 = \lim_{M \rightarrow \infty} \frac{E_0}{M} = \int d\chi \rho_{(1+)}(\chi) \epsilon_{(1+)}(\lambda) - 2 \sum_{k=1}^{\frac{N-1}{2}} \cot \left( \frac{\pi k}{N} \right) 
\eeq

Total energy is real since the imaginary part of the energy contribution from holes cancel 
the imaginary part of the energy of excitation, i.e., 
$ Im \left[ \epsilon_{(2s)} \right] = Im \left[ \epsilon_{(1+)} \right] $. The real parts are given by,
\beq
Re \left[ \epsilon_{(2s)} (\chi) \right] = Re \left[ \epsilon_{(1+)h } (\chi) \right]   
\eeq
In this case also, the dressed energy equals the bare energy. As before, we will denote by
$\epsilon_{(2s)} \left( \chi_{\beta}^{(2s)} \right)$  its real part.
The total energy is given by,
\beq
E = E_0 + \sum_{\beta = 1}^{M^{(2s)}} 2 \epsilon_{(2s)} \left( \chi_{\beta}^{(2s)} \right) 
\eeq
where $E_0$ is the ground state energy. \\

\noindent
We now turn to the calculation of momentum. 
The momentum associated with a string of length $j$ and
parity $v$ is found to be,
\beqar
p_{(1+)} (\chi) = - \frac{1}{2} \Theta_{(1+)}^{(1)} (\chi) \\
p_{(j)} (\chi) = - \frac{1}{2} \Theta_{(j)}^{(1)} (\chi ),  \hskip 1cm \mbox{for} \;\; j\neq (1+)
\eeqar
whence
\beq
p_{(2s)} (\chi) = 2 \arctan \left( \tanh \left( \frac{\chi}{2} \right) \right) 
\eeq

\noindent
If one keeps in mind the fact that the correlation of $\chi$s demand an additional $\pi$
for the creation of $(1+)$ hole, we can derive the dispersion relation

\beq
\epsilon_{(2s)} ( p ) = 4  \sin \left( \frac{p}{2} \right) 
\eeq

\section*{Acknowledgement}
Authors gratefully acknowledge the stimulating discussions with
Prof. Barry McCoy and Prof Leon Takhtajan.
Authors would like to thank the referee for his/her suggestion
to include details presented in sections II through IV in order 
to improve the clarity and readability of the paper.


\begin{thebibliography}{0}
\bibitem{pFZ1} V.A.Fateev and A.B.Zamolodchikov, Phys. Lett. {\bf 92A} (1982) 37.
\bibitem{pBBP} R.J.Baxter, V.V.Bazhanov and J.H.H.Perk, Int. J. Mod. Phys. {\bf B4} (1990) 803.
\bibitem{Sray01} Subhankar Ray, Ph.D. Dissertation, Institute for Theoretical Physics, Stony Brook, May 1994.
\bibitem{srjmp} Subhankar Ray, J. Math. Phys {\bf 38}(3) (1997) 1524.
\bibitem{pbethe} H.A.Bethe, Z. Phys. {\bf 71} (1931) 205.
\bibitem{albert1} G. Albertini, J. Phys. A {\bf 25} (1992) 1799.
\bibitem{srpl1} S. Ray, Phys. Lett. {\bf A 218} (1996) 80.
\bibitem{srpl2} S. Ray, Phys. Lett. {\bf A 222} (1996) 440.
\bibitem{AD} L.V.Avdeev and B.D.D\"{o}rfel, Nucl. Phys. {\bf B257 [FS14]} (1985) 253.
\bibitem{BdVV} O.Babelon, H.J.de Vega and C.M.Viallet, Nucl. Phys. {\bf B220 [FS8]} (1983) 13.
\bibitem{pkorebook} V. E. Korepin, N. M. Bogoliubov, A. G. Izergin, ``Quantum inverse scattering method and correlation functions'', Cambridge University Press (1993).
\bibitem{ptak71} M.Takahashi, Prog. Theo. Phys. {\bf 46} (1971) 401.
\bibitem{pkiri85} A.N. Kirillov, J. Sov. Math. {\bf 30} (1985) 2298.
\bibitem{pkiri87} A.N. Kirillov, J. Sov. Math. {\bf 36} (1987) 115.
\bibitem{pleon82} Leon Takhtajan, Phys. Lett. {\bf A 87} (1982) 479.
\bibitem{pKC} L. P. Kadanoff, H. Ceva, Phys. Rev. {\bf B3} (1971) 3918.
\bibitem{Bbook} R. J. Baxter, {\it ``Exactly Solvable Models in Statistical Mechanics''}, Academic Press (1982).
\bibitem{CoBook} R. J. Baxter, Exactly Solved Models in: {\it ``Fundamental
Problems in Statistical Mechanics''}, vol V, Ed. E. G. D. Cohen, North Holland,
Amsterdam (1980).
\bibitem{Za2} A. B. Zamolodchikov, Phys. Lett. {\bf B 97}, 63 (1980).
\bibitem{AK1} F. C. Alcaraz, R. Koberle, J. Phys. {\bf A 13}, L153 (1980).
\bibitem{YMPTY} H. Au-Yang, B. M. McCoy, J. H. H. Perk, S. Tang, M. L. Yan,
Phys. Lett. {\bf A 123}, 219 (1987).
\bibitem{MPTS} B. M. McCoy, J. H. H. Perk, S. Tang, C. H. Sah, Phys. Lett.
{\bf A 125}, 9 (1987).
\bibitem{BPY} R. J. Baxter, J. H. H. Perk, H. Au-Yang, Phys. Lett. {\bf A 128}, 138 (1988).
\bibitem{BS} V. V. Bazhanov, Yu G. Stroganov, J. Stat. Phys. {\bf 59}, 799 (1990).
\bibitem{ADM} G.Albertini, S.Dasmahapatra, B.M.McCoy, Phys.Lett {\bf A170}(1992), 397.
\bibitem{TS} M.Takahashi and M.Suzuki, Prog. Theor. Phys. {\bf 48} (1972) 2187.
\end{thebibliography}
\end{document}